\documentstyle[psfig,prb,aps,multicol]{revtex}

\draft
\begin{document}

\title{Elementary excitations in the gapped phase of a frustrated
  $\mathbf S=1/2$ spin ladder: \\ from spinons to the Haldane triplet}

\author{S. Brehmer, A. K. Kolezhuk,\cite{perm} H.-J. Mikeska, and
  U. Neugebauer} 
\address{Institut f\"ur Theoretische Physik, Universit\"at
  Hannover, Appelstr. 2, D-30167 Hannover, Germany}

\date{\today}
\maketitle

\begin{abstract}
  We use the variational matrix-product ansatz to study elementary
  excitations in the $S={1\over2}$ ladder with additional diagonal
  coupling, equivalent to a single $S={1\over2}$
  chain with alternating exchange and next-nearest neighbor
  interaction. In absence of alternation the elementary excitation
  consists of two free $S={1\over2}$ particles (``spinons'') which are
  solitons in the dimer order. When the nearest-neighbor exchange 
  alternates, the ``spinons'' are confined into one $S=1$ excitation
  being a soliton in the generalized string order. Variational results
  are found to be in a qualitative agreement with the exact
  diagonalization data for 24 spins.  We argue that such an approach
  gives a reasonably good description in a wide range of the model
  parameters.
\end{abstract}


\maketitle

\begin{multicols}{2}

\section{Introduction}
\label{sec:intro}
Spin ladders continue to attract much attention as structures intermediate
between one- and two-dimensional ones and possibly important for the
understanding of the high-T$_c$ superconductivity.  \cite{DagottoRice96} On
the other hand, there exists a close relationship between ``generalized'' spin
ladders (with an additional diagonal coupling), antiferromagnetic chains with
frustrating next-nearest neighbor interaction, and the Haldane systems.

\begin{figure}
\mbox{\hspace{20mm}\psfig{figure=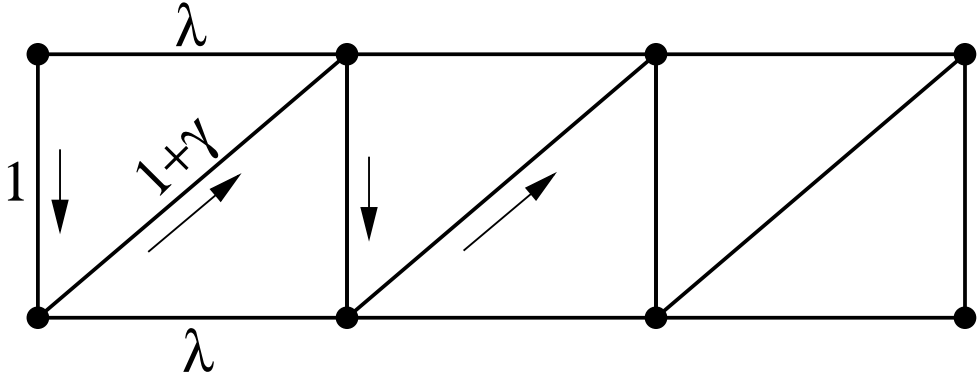,width=130mm,angle=-90.}}
  \caption{Generalized spin ladder with additional diagonal (frustrating)
    coupling. Arrows show the way of numbering the sites to map this system
    onto a single chain with nearest and next-nearest neighbor interactions.}
  \label{fig:gen-ladder}
\end{figure}

\begin{figure}
\mbox{\psfig{figure=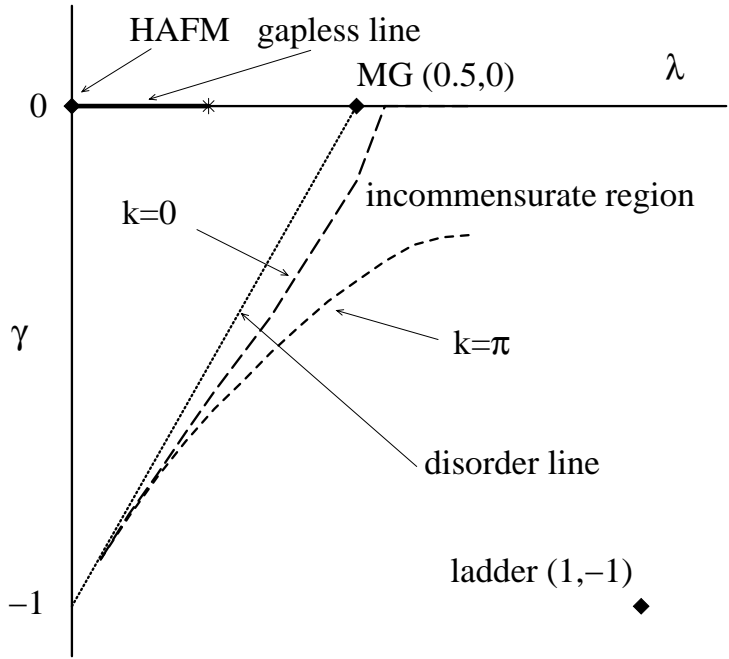,width=140mm,angle=-90.}}  \caption{Phase
  diagram of the generalized ladder. The interval of the line
  $\gamma=0$ for $0<\lambda<\lambda_{c}\approx 0.2411$ is gapless, and
  the other part of the diagram is gapped.  $k=0$ and $k=\pi$
  boundaries of the incommensurate region are variational estimates
  for the corresponding Lifshitz lines determined from the variation
  of the wave vector $k$ of the lowest excitation mode. The disorder
  line, where spin-spin correlations in the real space become
  incommensurate, for $k=0$ boundary is
  known\cite{TonegawaHarada87,Chitra+95} to be $\gamma=2\lambda-1$;
  for $k=\pi$ boundary the exact position of the disorder line is
  unknown. }
\label{fig:phasediag}
\end{figure}

In the present paper we study elementary excitations of the
generalized $S={1\over2}$ spin ladder model (equivalent to a single
zigzag spin chain with alternation and frustration). The model is
described by the Hamiltonian (see Fig.\ \ref{fig:gen-ladder})
\begin{eqnarray}
  \label{ham}
  \widehat H &=&\sum_{n} {\mathbf S}_{1,n}{\mathbf S}_{2,n}
  +(1+\gamma)\sum_{n} {\mathbf S}_{1,n} {\mathbf S}_{2,n+1}\nonumber\\ 
  &+&{\lambda}\sum_{n}({\mathbf S}_{1,n}{\mathbf S}_{1,n+1} +{\mathbf
    S}_{2,n}{\mathbf S}_{2,n+1})\;,
\end{eqnarray}
where $\lambda>0$ determines the strength of the next-nearest neighbor
interaction, and $\gamma$ corresponds to alternation of the
nearest-neighbor exchange, whose strength is set to be unity. This model
has rich behavior depending on the values of parameters $\lambda$ and
$\gamma$, and its phase diagram (see Fig. \ref{fig:phasediag}) is
rather well understood \cite{Chitra+95,Brehmer+96} (at least for the
half-plane $\lambda>0$; for negative $\lambda$ the situation is less
clear \cite{KolMik97}). Without loss of generality, we will assume
that $\gamma<0$, since there is an obvious symmetry
transformation\cite{Brehmer+96} relating the half-plane $\gamma>0$
with the strip $-1<\gamma<0$. The ``symmetry line'' $\gamma=0$ is
peculiar because it is the line of transition between dimerized phases
with different signs of the dimer order. Within the interval
$0<\lambda<\lambda_{c}\approx0.24$ this transition is of the second
order, the ground state is unique and nondimerized, and the
corresponding spectra are gapless; on the other part of this line the
transition is of the first order, so that for $\lambda>\lambda_{c}$,
$\gamma=0$ there are two degenerate dimerized ground states, and the
system is gapped. The transition at $\lambda=\lambda_{c}$ is well
studied, see Ref.\ \onlinecite{HaradaTonegawa93} for a review.
Everywhere except at the symmetry line the model (\ref{ham}) has unique
ground state with a finite gap above it. In the limit $\gamma\to
-\infty$ the diagonal spins form effective $S=1$ units, and the
system becomes equivalent to the $S=1$ Haldane chain, with the
effective coupling constant $(1+2\lambda)/4$.

   Elementary excitations of the generalized ladder, however, have
been studied to much less extent than its ground state properties. The
Heisenberg point ($\lambda=0$, $\gamma=0$) is exactly solvable by
means of the Bethe ansatz technique, and the elementary excitations
are pairs of noninteracting $S={1\over2}$ entities
(``spinons''). \cite{FaddeevTakhtajan81} The ground state contains the
Fermi sea of spinons, and the excitations are of the particle-hole
type.

It is also known that at the so-called Majumdar-Ghosh (MG) point
($\gamma=0$, $\lambda={1\over2}$), where the exact (twofold
degenerate) ground state is a simple product of singlet dimers,
\cite{MajumdarGhosh69} the elementary excitation can be approximately
constructed as a pair of unbound spins above the completely dimerized
state. \cite{ShastrySutherland81} The elementary excitation is
composed of two $S={1\over2}$ entities which are kinks in the dimer
order and resemble spinons in that they are ``almost free'' (i.e.,
form scattering states for most values of momenta) in case of 
unbroken translational invariance.  When nonzero alternation is
present, one may expect that those ``spinons'' are confined into a
single $S=1$ particle.

In the Haldane limit (infinite alternation) the system has 
long-range hidden (string) order,\cite{KennedyTasaki92} and the
elementary excitation is known
\cite{Knabe88,FathSolyom93,ElstnerMik94} to be a soliton in the string
order.  The concept of string order was generalized to spin ladders
\cite{TakadaWatanabe92,Nishiyama+95,White96} and it was shown that
several ladder models exhibit long-range generalized string
correlations. \cite{White96} It was argued \cite{White96} that the
gapped phase of the spin ladder is the same as the Haldane phase of
the effective $S=1$ chain.

We construct the variational ansatz for the wave function of the
elementary excitation in the form of a matrix product, using the
recently proposed \cite{Brehmer+96} matrix-product-states approach to
the description of the ground state properties of spin ladders. The
ground state in this approach has a built-in generalized string order,
and we construct the excitation as the $S=1$ composite particle being
a kink in the string order and consisting of two bound $S={1\over2}$
entities.  The wave function of the kink pair contains only one
variational parameter $\xi$ having the meaning of the average pair
size (the localization length).  We show that at the MG point $\xi$ is
infinite and our ansatz reduces to that of Shastry and Sutherland;
\cite{ShastrySutherland81} when moving away from the MG point, $\xi$
drops down very quickly to the value of about a few lattice
constants. At the ``regular'' ladder point ($\gamma=-1$, $\lambda=1$)
the dispersion relation as obtained from our ansatz agrees well with
the results of other authors using different technics (see Ref.\
\onlinecite{DagottoRice96} and references therein).  In the Haldane
limit $\xi$ goes to zero, and our wave function transforms into the
``crackion'' ansatz introduced by F\'ath and S\'olyom
\cite{FathSolyom93} (see also Ref.\ \onlinecite{NeugebMik96}) for the
description of the Haldane triplet in the Affleck-Kennedy-Lieb-Tasaki
(AKLT) model. We compare our variational results with the numerical
data obtained through the exact diagonalization of a finite (24 spins)
ladder system, and find a reasonable agreement between the two
approaches. We conclude that our simple variational ansatz allows us
to study analytically at a qualitative level the crossover from free
to strongly bound spinons, giving a reasonably good description in a
wide range of physical models.

The paper is organized as follows: in Sect.\ \ref{sec:ansatz} we
introduce our ansatz for the elementary excitation.  In Section
\ref{sec:results} we present results from the variational calculation
in comparison with numerical data. The elementary excitations in
different regions of the phase diagram are discussed.  Finally, Sect.\
\ref{sec:summary} contains concluding remarks.

\section{Two-spinon ``composite particle'' ansatz}
\label{sec:ansatz}  

Recently, \cite{Brehmer+96} a variational wave function for the
description of the ground state properties of generalized spin ladders
was proposed in the form of a matrix-product (MP) state.  The MP
representation was first discussed by Fannes {\em et al.}
\cite{Fannes+89} in an abstract manner and later by Kl\"umper {\em et
al.}\cite{Klumper+91-93} for the $S=1$ deformed VBS chain, and has
found since then numerous applications in exact and variational
calculations
\cite{NiggZitt96,GSu96,TotsukaSuzuki95,KolMikYam97,KolMik97}.  For
periodic boundary conditions, the trial wave function for the ground
state of the ladder consisting of $2N$ spins can be written as
\begin{eqnarray}
  \label{mp}
&&  |\Psi_{0}\rangle =\text{Tr}\,(g_{1} g_{2}\cdots g_{N})\,,\qquad
\text{where}\nonumber\\
&&  g_i(u,v) = u\,\Big(\widehat{\openone}\cdot |s\rangle_{i}\Big)
+v \sum_{\mu=0,\pm1}\widehat{\sigma}_{\mu} \cdot |t_{\mu}\rangle_{i} 
\end{eqnarray}
Here the elementary matrix $g_{i}$ is constructed from the singlet state
$|s\rangle_{i}$ and three triplet states $|t_{\mu}\rangle_{i}$ of the
ladder diagonals; $\sigma_{\mu}$ are the Pauli matrices in a spherical basis,
$\sigma_0=\sigma_z$, $\sigma_{\pm1}= \mp 1/\sqrt{2}\, (\sigma_x\pm
i\sigma_y)$.  The parameters $u$ and $v$ in case of absence of external
magnetic field can be chosen real and satisfy the normalization
condition $u^{2}+3v^{2}=1$.

The wave function (\ref{mp}) has the following remarkable properties:
\cite{Brehmer+96} (i) for arbitrary $u$ and $v$ it is a global singlet (see
also Ref.\ \onlinecite{KolMikYam97} for details); (ii) it has a built-in
generalized string order defined as ``diluted antiferromagnetic order'':
$|t_{+1}\rangle$ and $|t_{-1}\rangle$ should occur in perfect
antiferromagnetic sequence, arbitrarily diluted by $|s\rangle$'s and
$|t_{0}\rangle$'s; (iii) both degenerate dimer ground states at the
Majumdar-Ghosh point can be written in the above form, as well as the
``valence-bond'' AKLT state \cite{AKLT} which approximates the ground state in
the effective $S=1$ Haldane limit. The two dimer ground states at the MG point
correspond to $u=1$, $v=0$ and $u=v={1\over2}$ respectively, and the AKLT
state corresponds to zero singlet weight ($u=0$, $v=1/ \sqrt{3}$). The state
(\ref{mp}) has very short-ranged correlations and therefore cannot be
considered as a good approximation in the gapless region of the phase diagram;
we will thus restrict ourselves to the study of the gapped phase only.

The variational energy $E_{0}=\langle\Psi_{0}|\widehat{H}|\Psi_{0}\rangle$
calculated with the trial function (\ref{mp}) has at most two minima as a
function of $u,v$: one is always located at $u=1$, $v=0$ and corresponds to
singlets on the diagonal links, and the position of the other minimum
$(u_{0},v_{0})$ depends on $\lambda$ and $\gamma$; this latter minimum is
absent in certain region of the phase diagram. The two minima have equal
energies only at the MG point, and for any other choice of the model
parameters they are inequivalent.\cite{Brehmer+96}

\begin{figure}
\mbox{\hspace{20mm}\psfig{figure=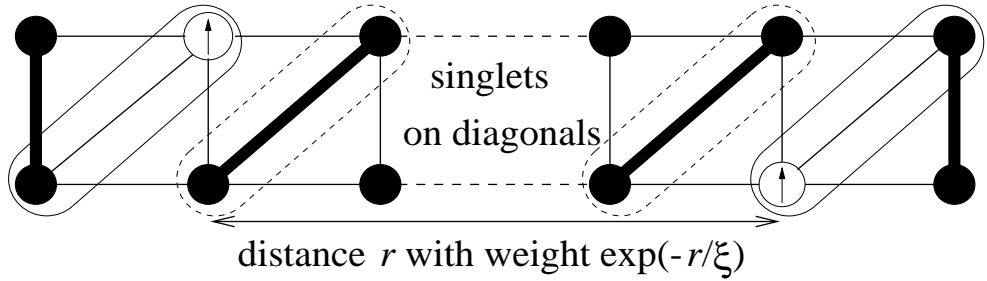,width=130mm,angle=-90.}}
\vspace{1mm} \caption{The structure of the variational two-spinon ansatz
  (\ref{ansatz1},\ref{ansatz2}). Ovals show the location of matrices;
  solid ovals denote $g^{(0)}$ and dashed ones denote
  $\widetilde{g}$. Thick solid links represent singlet bonds (bonds
  inside dashed ovals are {\em always} pure dimer bonds, and bonds
  between the solid ovals become purely singlet only on the disorder
  line $\gamma=2\lambda-1$).}  \label{fig:extendpic}
\end{figure}

We construct the trial wave function for the elementary excitation, requiring
the following: (i) it should be a triplet; (ii) it should be a soliton in the
generalized string order as defined above; (iii) it should be able to
reproduce the ansatz of Shastry and Sutherland for the MG point,
\cite{ShastrySutherland81} i.e.\ a pair of unbound spins connecting two
degenerate dimer ground states, and the ``crackion'' ansatz of F\'ath and
S\'olyom.\cite{FathSolyom93} One can check that the following construction
satisfies the above requirements: 
\begin{eqnarray}
\label{ansatz1}
&&|n,n';\mu\rangle=
\mbox{Tr}\Big\{ \Big(\prod_{i=1}^{n}{g}_{i}^{(0)}\Big)
\widehat \sigma_{\mu}^{\dagger}\prod_{i=n+1}^{n'} \widetilde{g}_{i} 
\prod_{i=n'+1}^{N} g_{i}^{(0)} \Big\}\,.
\end{eqnarray}
Here $g_{i}^{(0)}=g_{i}(u_{0},v_{0})$ denotes the matrix corresponding
to the variational ground state, and
$\widetilde{g}_{i}=g_{i}(u=1,v=0)$ is the matrix describing singlet
dimers on the diagonals (it is easy to see from (\ref{mp}) that
$\widetilde{g}_{i}$ is proportional to the unit matrix).  In fact, the
state (\ref{ansatz1}) describes two domain walls which correspond to
transitions between the two inequivalent variational ground states
mentioned above; it is worthwhile to remark that the states
$|n,n';\mu\rangle$ are not mutually orthogonal.  The presence of
$\widehat{\sigma}_{\mu}^{\dagger}$ ensures that this state breaks the
generalized string order and has the total spin $S=1$ and its
$z$-projection $S_{z}=\mu$ (the general technique of constructing MP
states with given quantum numbers $S$, $S_{z}$ can be found in Ref.\
\onlinecite{KolMikYam97}).  The structure of the ansatz
(\ref{ansatz1}) is schematically shown in Fig.\ \ref{fig:extendpic}.

 At the MG point $u_{0}=v_{0}={1\over2}$, and one can
straightforwardly check that in this case Eq.\ (\ref{ansatz1})
describes a pair of unbound spins $1\over2$ separating two completely
dimerized regions, i.e.\ it reduces to the ansatz of Shastry and
Sutherland.  On the other hand, if $n=n'$ and $u_{0}=0$, then the
state (\ref{ansatz1}) is exactly the same as the ``crackion'' of F\'ath
and S\'olyom.

We use the following (unnormalized) trial wave function of the composite
two-spinon excitation  with a given total
momentum $k$:
\begin{equation}
  \label{ansatz2}
  |k,q;\mu\rangle=\sum_{n'\geq n}e^{ik(n+n')/2}e^{iq(n-n')/2} e^{-(n'-n)/\xi}
  |n,n';\mu\rangle\,. 
\end{equation}
It contains two variational parameters $q$ and $\xi$ which can be considered
as real and imaginary part of the relative momentum of the two spin-$1\over2$
entities forming our composite particle. The parameter $\xi$ has the meaning
of a localization length, or the average size of the composite object, and
nonzero $q$ corresponds to the excitation of some internal degree of freedom.
If the localization length diverges, $\xi\to\infty$, the wave function
describes a triplet scattering state of two spinons, and finite
$\xi$ corresponds to a bound state.

On the disorder line $\gamma=2\lambda-1$, according to
Ref. \onlinecite{Brehmer+96}, the parameters $u_{0}$ and $v_{0}$ both equal
${1\over2}$, and the structure of our variational ansatz becomes rather
obvious: it describes a bound state of
two Shastry-Sutherland kinks (``free spins'' in a completely dimerized chain)
with the localization length $\xi$.

The energy for such an excitation can be calculated in the usual way:
\begin{equation}
  \label{disp}
E(q,\xi, k)={\langle k,q;\mu|\widehat{H}-E_{0}|k,q;\mu\rangle \over \langle
k;\mu|k;\mu\rangle }\;,
\end{equation}
and has to be minimized over $\xi$ and $q$, separately for any given $k$ (as
we will see below, it turns out that optimal $\xi$ strongly depends on
$k$). In this way one looks for a lowest variational state in a
subspace with the total spin $S=1$ and certain momentum $k$.   
Calculating the averages in (\ref{disp}) involving MP states can be
done with the help of the standard
technique. \cite{Klumper+91-93,TotsukaSuzuki95} The final expression for
$E(q,\xi,k)$ is quite cumbersome because of
the complicated structure of our trial wave function (\ref{ansatz1}), so that
we present here only the resulting dispersion plots for a number of
representative points of the phase diagram. The minimization has been
performed numerically.

Another, simpler ansatz can be obtained if one forces (\ref{ansatz2}) to be
strongly localized, i.e. $\xi\to0$. Then only the configurations with $n=n'$
survive, and we obtain
\begin{eqnarray}
  \label{ansatz3}
 && |k;\mu\rangle=\sum_{n}e^{ikn} |n;\mu\rangle\,,\;\; {\rm where} \nonumber\\ 
 &&|n;\mu\rangle= \mbox{Tr}\,\{ (g^{(0)}_{1}  g^{(0)}_{2} \cdots 
 g^{(0)}_{n})\widehat\sigma_{\mu}^{\dagger}(g^{(0)}_{n+1} g^{(0)}_{n+2} \cdots 
 g^{(0)}_{N} \}\,.
\end{eqnarray}
Such an ansatz may be called ``regular,'' or localized crackion (in
contrast to our two-spinon ansatz (\ref{ansatz2}) describing an
``extended crackion'') because it exactly reproduces the structure of
the wave function proposed by F\'ath and S\'olyom, and the only
difference is in the generalized concept of the string order (or, in
other words, in the fact that matrices ${g}_{i}^{(0)}$ are allowed to
contain singlet states). This ``regular'' crackion ansatz is
essentially equivalent to the construction introduced recently by
Nakamura {\em et al.}  \cite{Nakamura97} within a slightly different
approach using the Kennedy-Tasaki unitary transformation.

\section{Variational results for the excitations and comparison with
        numerical data}
\label{sec:results}

In this section we present variational and numerical results for
dispersion relations, lowest modes and their corresponding wave
vectors for various points of the phase diagram. In the first
subsection we study the spinon-type excitations on the symmetry line
$\gamma=0$ in the vicinity of the MG point.  Variational energies for
scattering states and bound states are computed at the MG point
$\lambda={1\over2}$, and the difference in behavior of the spectrum at
$\lambda>{1\over2}$ and $\lambda<{1\over2}$ is discussed. In the
second subsection we investigate the consequences of alternation.
When one moves off the symmetry line, the crossover from spinon-type
excitations to ladder-type (``crackion'') excitations occurs, being
characterized by the change of the localization length $\xi(k_{0})$
from infinity to zero; here $k_{0}$ denotes the wave vector of the
lowest energy mode. We show that this crossover takes place at the
$k_{0}=\pi$ boundary of the incommensurate region where $k_{0}$
changes gradually from $0$ to $\pi$. These results are compared with
numerical data from exact diagonalization for 24 spins. Because the
correlation lengths for the points taken into account are rather
small, the numerical data are very close to the thermodynamic limit
and therefore we do not perform any finite-size extrapolation.

\subsection{Elementary excitations on the symmetry line
  \boldmath$\gamma=0$\unboldmath} 

\begin{figure}
\mbox{\psfig{figure=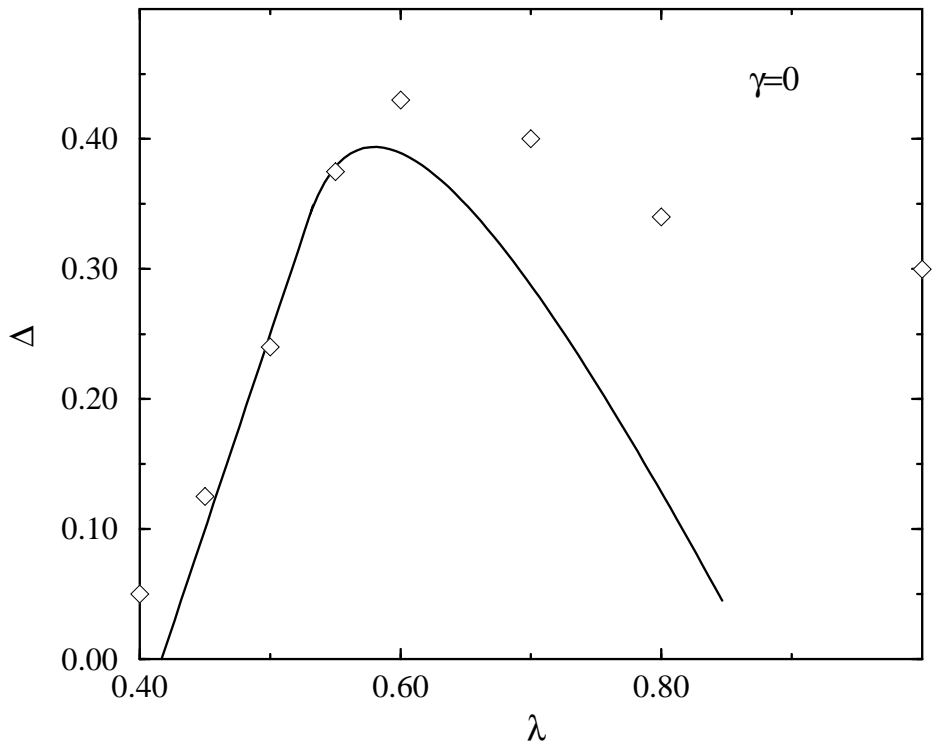,width=140mm,angle=-90.}}
    \caption{Dependence of the gap on $\lambda$ on the symmetry line $\gamma=0$
    in the vicinity of the MG point; full line represents the variational
    result according to (\ref{SS}), and squares are numerical (DMRG) results
    by White and Affleck.\protect\cite{WhiteAff96}}
  \label{fig:symgap}
\end{figure}

The ansatz (\ref{ansatz1},\ref{ansatz2}) obviously becomes inadequate
close to the symmetry line $\gamma=0$ (except for the MG point),
because two variational minima of the ground state energy are
inequivalent while the true ground state is twofold degenerate on this
line. However, the interval of the symmetry line in the vicinity of
the MG point (i.e.\ $\gamma=0$ and $\lambda$ close to ${1\over2}$) can
be studied with the help of the Shastry-Sutherland-type ansatz: it is
sufficient to put $u_{0}=v_{0}=1/2$ in
(\ref{ansatz1},\ref{ansatz2}). In the limit $\xi \rightarrow \infty$
one gets exactly the ansatz of Ref.\ \onlinecite{ShastrySutherland81}
and can calculate variational energies for the scattering states. On
the other hand, if we do not force $\xi = \infty$, the energy of the
lowest bound state can be calculated for each value of the total
momentum $k$. For the scattering states it is possible to obtain a
compact analytical expression for the energy of such a two-particle
excitation:
\begin{eqnarray}
 \label{SS}
 && E(k,q)=\varepsilon\Big({k+q\over2}\Big) +\varepsilon\Big({k-q\over2}\Big)
 \,,\\
 && \varepsilon(k)={\lambda\over4}(5+4\cos k)+(1-2\lambda)\Bigg\{
{3\over8} +{5\cos k+4\over 5+4\cos k}\Bigg\}\,, \nonumber
\end{eqnarray}
here $\varepsilon(k)$ has the meaning of the spinon dispersion.  At
$\lambda={1\over2}$ it coincides with the result of Ref.\
\onlinecite{ShastrySutherland81}.  (It should be mentioned that since
in this paper we consider the general case of an alternated chain,
momenta in Eqs.\ (\ref{ansatz2},\ref{SS}) are defined in the halved
Brillouin zone, in contrast to Ref.\
\onlinecite{ShastrySutherland81}). The dispersion relation is
determined by the lower boundary of the two-particle continuum. The
resulting  $\lambda$-dependence of the gap $\Delta$ is shown in Fig.\
\ref{fig:symgap}, together with the numerical results by White and
Affleck.\cite{WhiteAff96}  One can see that quantitatively this approach
yields reasonable results only in the close vicinity of the MG point;
nevertheless, at the qualitative level it correctly predicts closing of the
gap when decreasing $\lambda$ and the existence of a maximum in 
$E_{g}(\lambda)$ at $\lambda$ slightly greater than ${1\over2}$.

\begin{figure}
\mbox{\psfig{figure=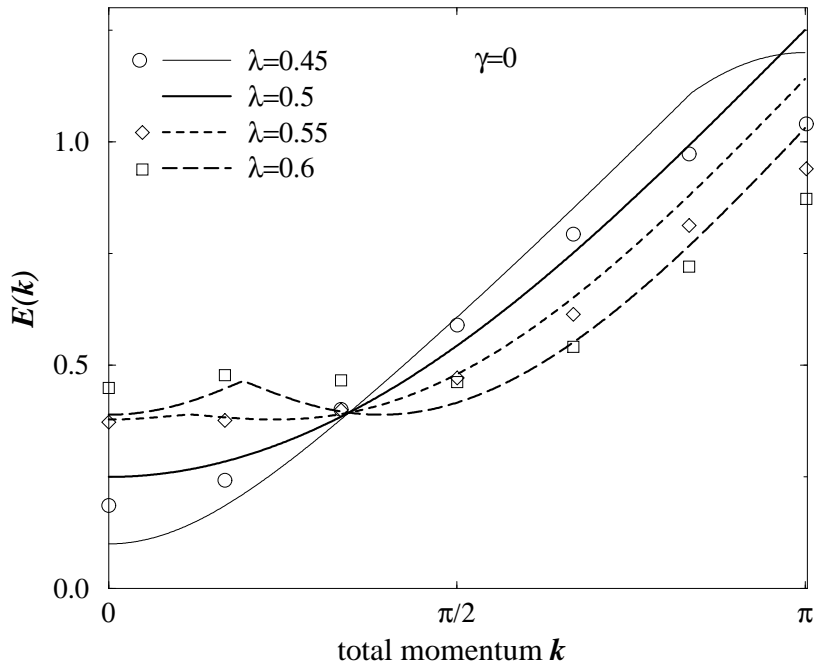,width=125mm,angle=-90.}}
    \caption{Typical plots of the lower boundary of the two-spinon continuum
      on the symmetry line $\gamma=0$
    in the vicinity of the MG point: lines correspond to the formula
    (\ref{SS}) and points ($\Box,\diamond,\circ$) are exact
    diagonalization data for a 24-spin system.}
  \label{fig:symspectra}
\end{figure}

 Characteristic plots of the dispersion of scattering states in the
vicinity of the MG point are shown in Fig.\ \ref{fig:symspectra}. From
(\ref{SS}) one can see that for $\lambda$ less than
$\lambda_{\pi}={9\over17}$ the single-spinon dispersion has a minimum
at $k=k_{0}=\pi$, and for larger $\lambda$ this minimum shifts towards
$k_{0}$ lower than $\pi$. Thus, the lower boundary of the two-spinon
continuum always has the minimum at $k=0$ (corresponding to
$q=2k_{0}$), and for $\lambda>\lambda_{\pi}$ there appears another
minimum at $k=k_{0}=2(\pi-k_{0})$ (which corresponds to $q=0$). When
$\lambda$ increases, this second minimum gets more pronounced.

Appearance of the lowest mode with incommensurate wave vector is
closely related to the existence of the so-called disorder points
\cite{Stephenson69,Garel86,Scholl96a} where spin-spin correlations in
real space become incommensurate. Strictly speaking, the point where
the wave vector of the lowest mode becomes incommensurate corresponds
not to the disorder point itself, but to the so-called Lifshitz point
where the correlation function peak in momentum space (i.e., peak in
the structure factor $S(q)$) starts moving from commensurate to
incommensurate $q$.  Generally, the Lifshitz point does not coincide
with the disorder point and is situated at some small distance from
the boundary of the incommensurate region.  In our variational
calculation the wave vector starts to change at
$\lambda_{\pi}={9\over17}$, while the disorder line (line of disorder
points) is numerically established \cite{TonegawaHarada87,Chitra+95}
to be $\gamma=2\lambda-1$, which means that at $\gamma=0$ the disorder
point is $\lambda={1\over2}<\lambda_{\pi}$, in agreement with the
above.

\begin{figure}
\mbox{\psfig{figure=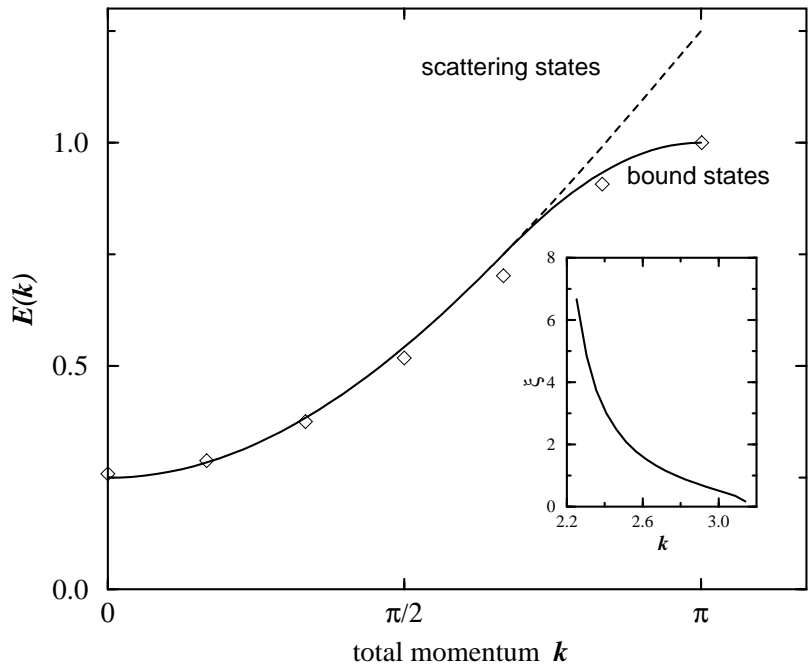,width=125mm,angle=-90.}}
  \caption{Variational result for the dispersion at the MG point
  $\gamma=0$, $\lambda={1\over2}$ (solid line) in comparison with the
  numerical data  for 24 spins (diamonds); the insert shows the momentum
  dependence of the localization length $\xi(k)$ for the bound
  states.}  \label{fig:MGdisp}
\end{figure}

We would like to end this subsection by pointing out the role of bound
states.  At the MG point it is known \cite{ShastrySutherland81} that
bound states are lower in energy than scattering states for wave
vectors $k$ close to the zone boundary. This was obtained in Ref.\
\onlinecite{ShastrySutherland81} by using the variational estimate for
the upper bound of the dispersion (see below).  We can capture the
dispersion of the lowest bound state in our approach if we minimize
with respect to $\xi$ for each $k$. For values of the total momentum $
0.68\pi < k \le \pi $ we obtain $ 1/ \xi_{\rm min} > 0$ as shown in
Fig.\ \ref{fig:MGdisp}. The same feature, namely the appearance of
bound states as the lowest energy excitations, can be expected for
$\lambda>\lambda_{\pi}$ and $k$ around the midpoint in between the
two dispersion minima, but we did not perform this calculation.
Generally, more than one bound state may exist; \cite{Kampf+}
unfortunately, within the present approach one can access only the
{\em lowest} bound state.

\subsection{Dispersion relations and crossover from loosely bound to
tightly bound spinons}

First of all, we would like to focus on the disorder line
$\gamma=2\lambda-1$. As soon as one moves off the MG point
($\lambda={1\over2}$, $\gamma=0$) towards the dimer point
($\lambda=0$, $\gamma=-1$), the energy of the MG ground state with
singlets on diagonals picks up an energy proportional to the size of
the system,
while the other MG ground state with singlets on the rungs remains the
true ground state.  As a consequence, only bound states of spinons can
survive: this is a typical confinement situation.  The wave vector of
the lowest mode is still $k=0$.

\begin{figure}
\mbox{\psfig{figure=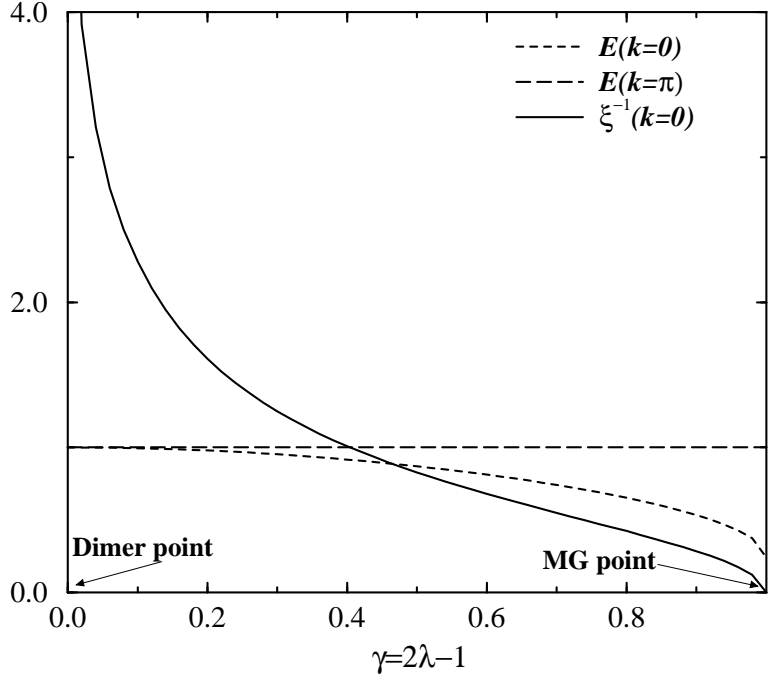,width=125mm,angle=-90.}}
  \caption{The bandwidth $E(0)-E(\pi)$ and the localization 
    length $\xi$ of the lowest energy mode on the disorder line
 $\gamma=2\lambda-1$.}
  \label{fig:disord}
\end{figure}

The main feature of this excitation is the $k$-dependent
delocalization: The ground state is formed by a product of rung
singlets. If we now replace one singlet by a triplet and superpose
with wave vector $\pi$, we obtain an exact eigenstate on the disorder
line, whose energy gives the upper bound of the
dispersion. \cite{CaspersMagnus}  To obtain the lowest mode we delocalize
the up spins, which make up the triplet, with the amplitude $e^{-\xi /
r}$ as shown in Fig. \ \ref{fig:extendpic} and superpose with the wave
vector $k=0$.  The bandwidth for this excitation is illustrated in
Fig. \ \ref{fig:disord}; one can see that the bandwidth increases
along with the increase of the localization length $\xi$ from $0$ to
$\infty$ on the way from the dimer point to the MG point (here $\xi$
is not to be confused with the spin correlation length!). At the MG
point we reproduce the result of Shastry and Sutherland for the gap
$\Delta=1/4$.  One may say that here we observe the crossover from
loosely bound spinons (``extended crackion'', $1 \ll \xi < \infty$) to
spinons tightly bound into the Haldane triplet (``crackion,'' $\xi \ll
1$), even though the crackion at the dimer point is a trivial
(dispersionless) excitation.

The same physical picture of crossover should be valid for any path beginning
somewhere at the symmetry line and ending somewhere at sufficiently negative
$\gamma$; on the symmetry line $\xi(k_{0})$ should be infinite ($k_{0}$
denotes the lowest mode wave vector), and it decreases to zero when
$-\gamma$ is large enough.  In the present approach we can observe this
$\xi\to\infty$ behavior only for the MG point, because for any other point on
the symmetry line our two variational ground states become inequivalent and we
lose the feature of twofold degeneracy. However, we believe that our approach
remains reasonable for points which are far enough from the symmetry line,
where the translational symmetry is explicitly broken and this effect
overrides the built-in dimerization of our variational ansatz.

It is worthwhile to make a few remarks concerning the behavior of the real
part of relative momentum $q$. Because of the confinement, the energy of
states with internal motion, i.e. with $q\not=0,2\pi$, has to be much higher
compared to the states with zero relative momentum. In our variational
calculations we were able to detect only one minimum which always occured at
$q=0$ or $2\pi$ (strictly speaking, $q=0$ for $\pi<k<2\pi$ and $q=2\pi$ for
$0<k<\pi$, so that sets with $q=0$ and $q=2\pi$ correspond to physically
equivalent states).

\begin{figure}
\mbox{\psfig{figure=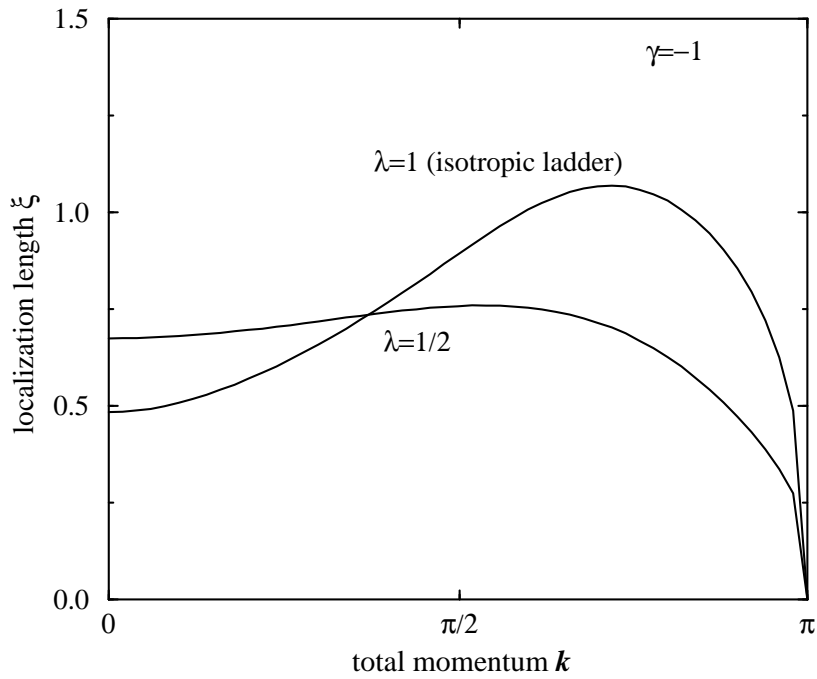,width=125mm,angle=-90.}}
  \caption{Momentum dependence of the localization 
    length $\xi$ for two points on the line $\gamma=-1$. Note that
 $\xi(\pi)=0$.}
  \label{fig:gamma-1xi}
\end{figure}

\begin{figure}
\mbox{\psfig{figure=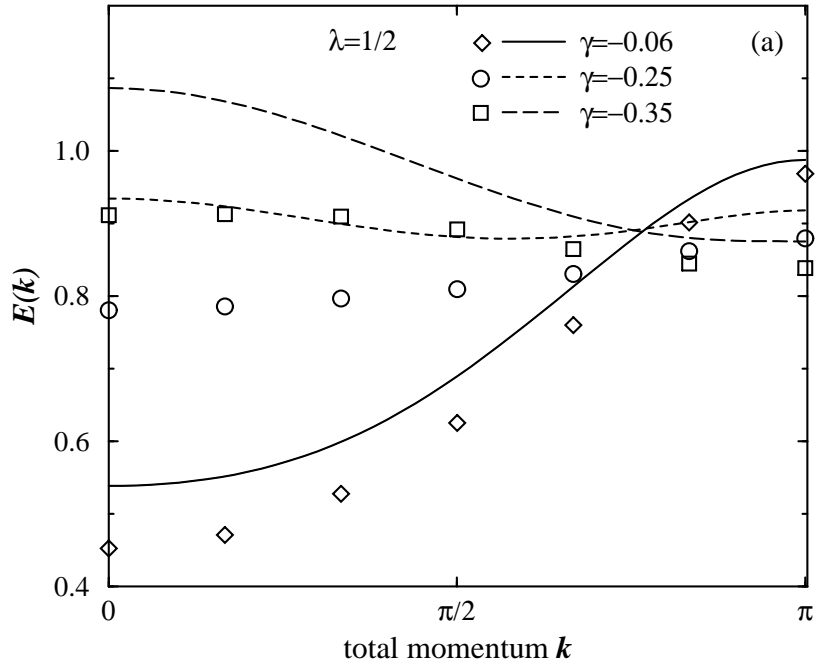,width=125mm,angle=-90.}} \vskip 2mm
\mbox{\psfig{figure=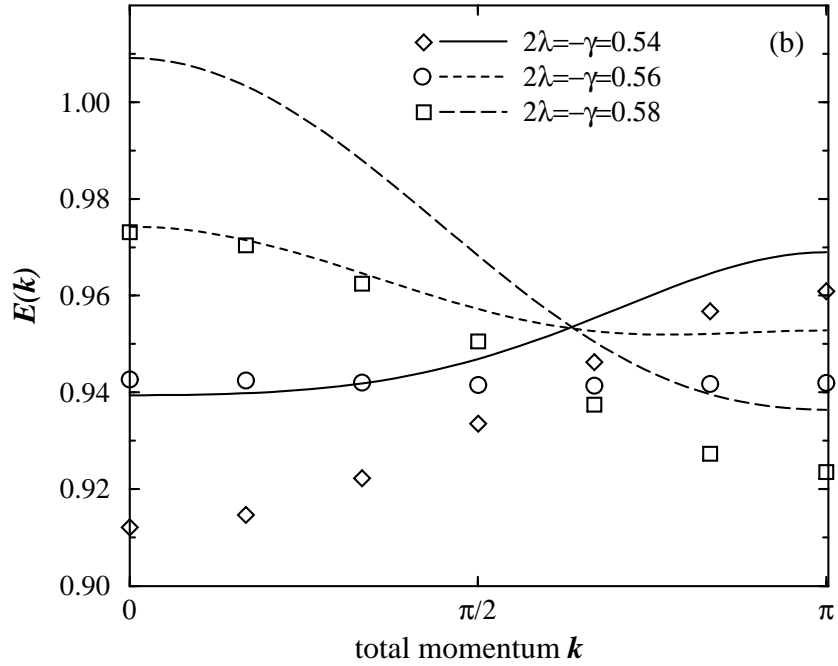,width=125mm,angle=-90.}}  
\caption{Dispersion curves from variational ansatz in comparison with
    the numerical data from exact diagonalization for 24 spins: (a)
    three points on the line $\lambda={1\over2}$; (b) three points on
    the line $\gamma=-2\lambda$ in the vicinity of its crossing with
    the disorder line $\gamma=2\lambda-1$ (numerical data are taken
    from Ref.\ \protect\onlinecite{Brehmer+96}).}  \label{fig:disp}
\end{figure}

The crossover from `extended crackion' to crackion is related to the
change of wave vector $k_{0}$ of the lowest mode, due to the following
remarkable feature which we observed from our calculations and which
is illustrated by Fig.\ \ref{fig:gamma-1xi}: {\em everywhere in the
gapped region of the phase diagram the property $\lim_{k \to \pi}
\xi(k) = 0$ holds.}  Thus, once $k_{0}$ has changed from $0$ to $\pi$,
we know that the lowest mode is a `usual' crackion because $\xi(
\pi)=0$, so that the crossover takes place at the $k_{0}=\pi$ boundary
of the incommensurate region.  In order to determine the boundaries of
the incommensurate region, we used our variational approach and
compared the results with exact diagonalization data.  In Fig.\
\ref{fig:disp}(a) dispersion relations for a few points on the
vertical line $\lambda={1\over2}$ are presented. Numerical data and
variational results are in a good agreement even though the lowest
wave vectors are at slightly different positions (one should keep in
mind that numerical dispersions for finite chains consist of a finite
number of points). Fig.\ \ref{fig:disp}(b) illustrates the change of
$k_{0}$ when crossing the disorder line.  Similarly to the situation
on the symmetry line as described in the previous subsection, one can
see that $k_{0}$ starts to change from $0$ not exactly at the disorder
line but slightly above it (i.e., the Lifshitz line is not identical
to the disorder line). The comparison with the numerical data as is
shown in Fig.\ \ref{fig:disp}(a) confirms this property, within the
numerical accuracy (exact diagonalization of 24 spins leads to 12
values for the wave vector which is not sufficient to mark the
incommensurate region precisely but allows qualitative comparison).
The boundaries of the incommensurate region obtained by the
variational calculation are presented in Fig.\ \ref{fig:phasediag}. It
should be pointed out that our result for the $k_{0}=\pi$ boundary
does not agree with that of Pati et al. \cite{Pati+96} For example, we
obtain that the $k_{0}=\pi$ boundary goes through the dimer point
$\gamma=-1$, $\lambda=0$, while the corresponding curve $C$ of Fig.\ 2
from Ref.\ \onlinecite{Pati+96} crosses the $\gamma=-1$ line at
$\lambda\approx0.6$. At present we cannot comment on the origin of
this strong discrepancy.

\begin{figure}
\mbox{\psfig{figure=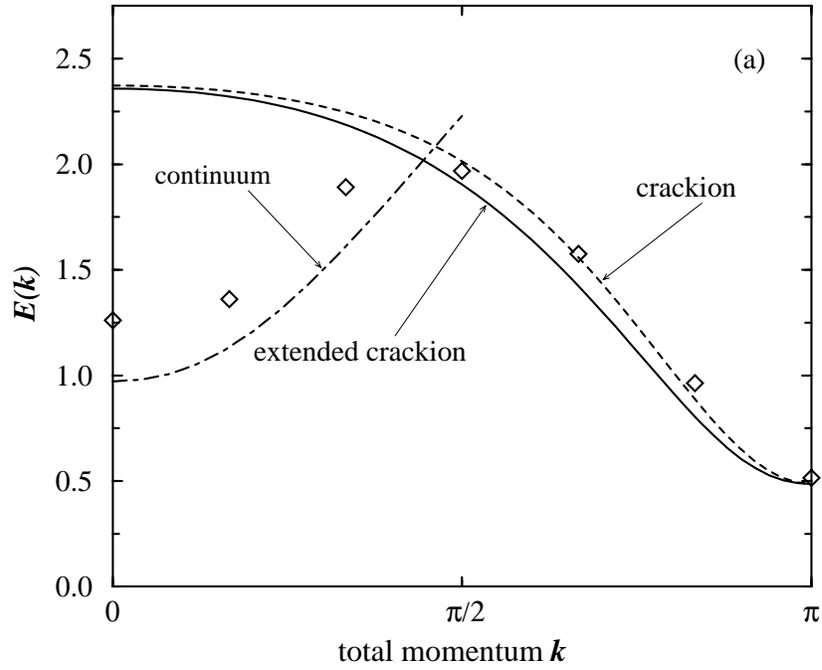,width=125mm,angle=-90.}}
\vskip 2mm \mbox{\psfig{figure=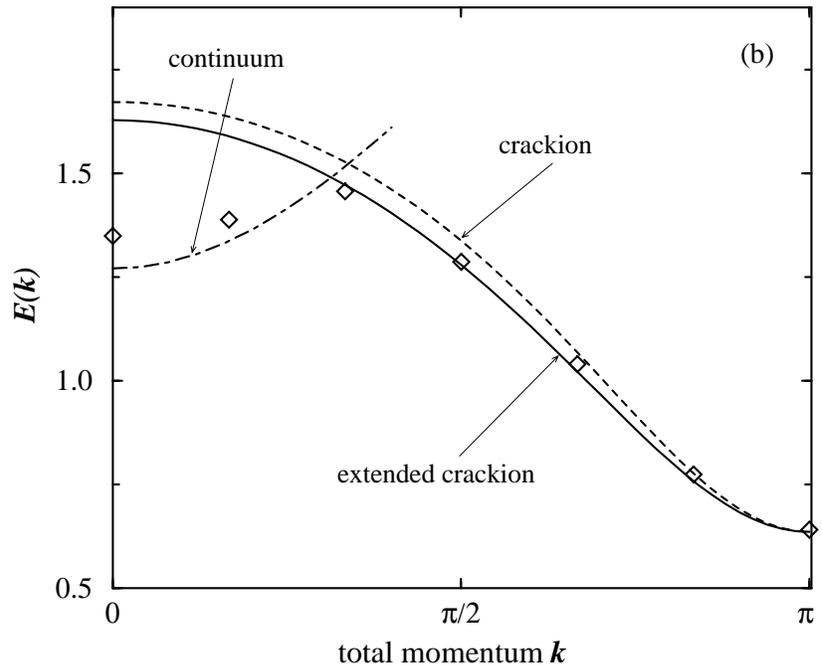,width=125mm,angle=-90.}}
\caption{The dispersion of elementary excitations on the
line $\gamma=-1$: (a) the ladder point $\lambda=1$; (b)
$\lambda={1\over2}$. Variational results from ``extended crackion''
($\xi=\xi_{\rm min}(k)$) and ``crackion'' ($\xi$=0) approaches are
shown along with the numerical data ($\Diamond$) from exact
diagonalization. The dash-dotted curve shows the lower boundary of the
two-particle continuum for `extended crackion' ansatz.}
\label{fig:dispgamma-1}
\end{figure}

\begin{figure}
\mbox{\psfig{figure=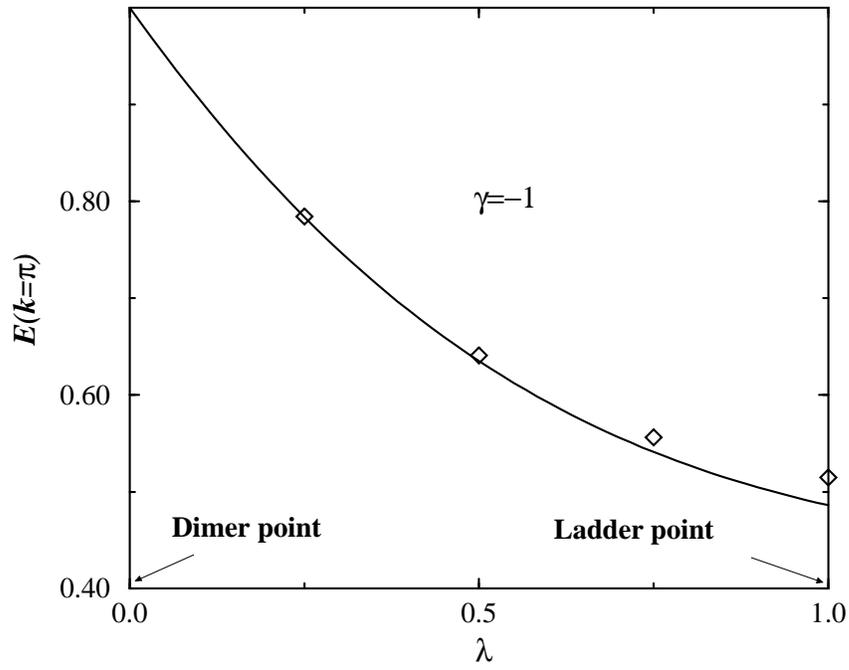,width=125mm,angle=-90.}}
  \caption{The energy of the lowest mode $E(k=\pi)$ along the line $\gamma=-1$;
    solid line represents the variational result and diamonds ($\Diamond$) are
    numerical points from exact diagonalization for 24 spins.}
  \label{fig:LMgamma-1}
\end{figure}

Finally, we would like to discuss the vicinity of the line $\gamma=-1$
which includes the experimentally relevant ladder point ($\lambda=1$,
$\gamma=-1$). Figure \ref{fig:dispgamma-1} shows the dispersion curves
for the ``regular'' crackion, the ``extended'' crackion and the
numerical data for two points on this line (one should mention that
our data agree rather well with those of
Ref. \onlinecite{BarnesRiera94}).  The dispersion curve of the
extended crackion is located slightly below the curve of the crackion
and coincides with it for $k= \pi$, in agreement with the general
property $\lim_{k \to \pi} \xi(k) = 0$ (see also Fig. \
\ref{fig:gamma-1xi}).  One can also observe that the $k=0$ gap is
slightly larger than $2E(k=\pi)$, which indicates the repulsive
character of the effective interaction between the elementary
excitations in the ladder.  Fig. \ \ref{fig:LMgamma-1} displays the
lowest crackion mode $(k= \pi)$ in comparison with exact
diagonalization data along the horizontal line $\gamma=-1$, $\lambda=
0 \dots 1$.  We conclude that the ladder excitations can be described
by the usual crackion ansatz, and the isotropic spin ladder has the
same type of elementary excitation as the effective $S=1$ chain which
appears as the limit $\gamma \to - \infty$ of the generalized
frustrated ladder.  In that limit the localization length $\xi$
collapses for all values of $k$ because creation of singlets on the
diagonals would cost infinite energy.

\section{Conclusion}
\label{sec:summary}

We have presented a variational matrix-product ansatz for elementary
excitations in the gapped phase of the $S={1\over2}$ ladder with
an additional frustrating diagonal coupling $1+\gamma$, $\gamma<0$; the
strength of interaction along the legs is $\lambda$, and the interaction
along the rungs is chosen to be unity. This system is equivalent to the
antiferromagnetic spin-${1\over2}$ zigzag chain with alternating exchange (the
magnitude of alternation is proportional to $\gamma$) and next-nearest
neighbor interaction $\lambda$. Our ansatz describes a triplet state of two
$S={1\over2}$ entities (``spinons'') and allows one to interpolate between 
free and bound spinons by varying the parameter $\xi$
which has the meaning of a localization length (average distance between
spinons in the pair). This state is constructed to be a soliton in generalized
string order \cite{Brehmer+96} and in the limit $\gamma\to-\infty$ of
effective $S=1$ Haldane chain it coincides with the ``crackion'' ansatz
proposed by F\'ath and S\'olyom \cite{FathSolyom93} if the localization length
$\xi\to0$; for that reason we call our ansatz an ``extended crackion.''  The
limit $\xi \to \infty$ leads to the two-particle excitation of Shastry and
Sutherland, \cite{ShastrySutherland81} which corresponds to free spinons
existing in absence of the alternation (i.e., on the symmetry line
$\gamma=0$).

Using our variational ansatz, we calculated dispersion relations for
various points in the phase diagram. These results were compared to
exact numerical diagonalization data for 24 spins and showed a
reasonable agreement of the two approaches. The variational parameter
$\xi$ was determined separately for each value of the total momentum
$k$; it turns out that $\xi(k)$ has nontrivial behavior, particularly
the property $\lim_{k\to\pi}\xi=0$ was numerically observed in the
entire range of studied model parameters.

We determined the boundaries of the incommensurate region (strictly speaking,
the corresponding Lifshitz lines) by locating the wave vector $k_{0}$ of the
lowest mode:  for $\gamma>2\lambda-1$ the wave vector is
pinned at $k_{0}=0$, slightly after crossing this line it starts to change
from $0$ to $\pi$, and, finally, at some other line it gets again pinned at
$k_{0}=\pi$ (in terms of the full Brillouin zone of the chain this corresponds
to the change from $\pi$ to $\pi/2$). 

We show that in the interval between the symmetry line $\gamma=0$ and
the $k_{0} = \pi$ boundary of the incommensurate region the dispersion
of elementary excitations is well described by our bound spinons
ansatz.  The crossover of the lowest mode from ``extended crackion'' to
``localized'' crackion (i.e., from finite $\xi(k_{0})$ to
$\xi(k_{0})=0$) occurs at the $k_{0}=\pi$ boundary of the
incommensurate region, due to the above-mentioned property of the
function $\xi(k)$. For the isotropic ladder point ($\gamma=-1$,
$\lambda=1$) a localized crackion ansatz with $\xi=0$ is sufficient to
describe the excitations.

To conclude, we propose a simple ansatz providing a reasonably good
description of the elementary excitations in the gapped phase of the
frustrated $S={1\over2}$ chain with alternation for a wide range of the
model parameters.

\acknowledgements

 This work was supported by the German Federal Ministry for Research
and Technology (BMBF) under the contract 03MI4HAN8.  One of us (A.K.)
gratefully acknowledges the hospitality of Hannover Institute for
Theoretical Physics and the support by the Ukrainian Ministry of
Science (grant 2.4/27) and by Deutsche Forschungsgemeinschaft.

\end{multicols}

\newpage

\end{document}